\documentclass[12pt,axodraw]{article}
\usepackage[sumlimits]{amsmath}
\usepackage{amsmath}
\usepackage{epsfig}
\usepackage{amssymb}
\usepackage{graphicx}
\usepackage{axodraw}
\usepackage{pstricks}
\usepackage{color}

\newcommand{\be}{\begin{equation}}
\newcommand{\ee}{\end{equation}}
\newcommand{\beq}{\begin{equation}}
\newcommand{\eeq}{\end{equation}}
\newcommand{\ba}{\begin{eqnarray}}
\newcommand{\ea}{\end{eqnarray}}
\newcommand{\bea}{\begin{eqnarray}}
\newcommand{\eea}{\end{eqnarray}}

\begin{document}
\pagestyle{plain} \setcounter{page}{1}


\def\del{{\partial}}
\def\vev#1{\left\langle #1 \right\rangle}
\def\cn{{\cal N}}
\def\co{{\cal O}}
\def\IC{{\mathbb C}}
\def\IR{{\mathbb R}}
\def\IZ{{\mathbb Z}}
\def\RP{{\bf RP}}
\def\CP{{\bf CP}}
\def\Poincare{{Poincar\`e}}
\def\tr{{\rm tr}}
\def\tp{{\tilde \Phi}}
\def\TL{\hfil$\displaystyle{##}$}
\def\TR{$\displaystyle{{}##}$\hfil}
\def\TC{\hfil$\displaystyle{##}$\hfil}
\def\TT{\hbox{##}}
\def\HLINE{\noalign{\vskip1\jot}\hline\noalign{\vskip1\jot}} 
\def\seqalign#1#2{\vcenter{\openup1\jot
  \halign{\strut #1\cr #2 \cr}}}
\def\lbldef#1#2{\expandafter\gdef\csname #1\endcsname {#2}}
\def\eqn#1#2{\lbldef{#1}{(\ref{#1})}%
\begin{equation} #2 \label{#1} \end{equation}}
\def\eqalign#1{\vcenter{\openup1\jot
    \halign{\strut\span\TL & \span\TR\cr #1 \cr
   }}}
\def\eno#1{(\ref{#1})}
\def\href#1#2{#2}
\def\half{{1 \over 2}}

\def\ads{{\it AdS}}
\def\adsp{{\it AdS}$_{p+2}$}
\def\cft{{\it CFT}}

\newcommand{\gsim}{\lower.7ex\hbox{$\;\stackrel{\textstyle>}{\sim}\;$}}
\newcommand{\lsim}{\lower.7ex\hbox{$\;\stackrel{\textstyle<}{\sim}\;$}}
\newcommand{\ber}{\begin{eqnarray}}
\newcommand{\eer}{\end{eqnarray}}
\newcommand{\beqar}{\begin{eqnarray}}
\newcommand{\cN}{{\cal N}}
\newcommand{\cO}{{\cal O}}
\newcommand{\cA}{{\cal A}}
\newcommand{\cT}{{\cal T}}
\newcommand{\cF}{{\cal F}}
\newcommand{\cC}{{\cal C}}
\newcommand{\cR}{{\cal R}}
\newcommand{\cW}{{\cal W}}
\newcommand{\eeqar}{\end{eqnarray}}
\newcommand{\eps}{\epsilon}
\newcommand{\pa}{\paragraph}
\newcommand{\pt}{\partial}
\newcommand{\de}{\delta}
\newcommand{\De}{\Delta}
\newcommand{\lb}{\label}


\newcommand{\oh}{\displaystyle{\frac{1}{2}}}
\newcommand{\dsl}
  {\kern.06em\hbox{\raise.15ex\hbox{$/$}\kern-.56em\hbox{$\partial$}}}
\newcommand{\id}{i\!\!\not\!\partial}
\newcommand{\as}{\not\!\! A}
\newcommand{\ps}{\not\! p}
\newcommand{\ks}{\not\! k}
\newcommand{\D}{{\cal{D}}}
\newcommand{\dv}{d^2x}
\newcommand{\Z}{{\cal Z}}
\newcommand{\N}{{\cal N}}
\newcommand{\Dsl}{\not\!\! D}
\newcommand{\Bsl}{\not\!\! B}
\newcommand{\Psl}{\not\!\! P}
\newcommand{\ZZ}{{\rm \kern 0.275em Z \kern -0.92em Z}\;}


\newcommand{\lbl}[1]{\label{eq:#1}}
\newcommand{ \rf}[1]{(\ref{eq:#1})}
\newcommand{\setl}{\setlength\arraycolsep{2pt}}

\newcommand{\noi}{\noindent}
\newcommand{\ra}{\rightarrow}
\newcommand{\Ra}{\Rightarrow}

\newcommand{\cd}{\bar{D}}
\newcommand{\cc}{\bar{C}}
\newcommand{\cB}{{\cal B}}
\newcommand{\cD}{{\cal D}}
\newcommand{\cG}{{\cal G}}
\newcommand{\cH}{{\cal H}}
\newcommand{\cK}{{\cal K}}
\newcommand{\cL}{{\cal L}}
\newcommand{\cM}{{\cal M}}
\newcommand{\cP}{{\cal P}}
\newcommand{\cS}{{\cal S}}
\newcommand{\cU}{{\cal U}}

\newcommand{\CF}{C_{\rm F}}
\newcommand{\Imm}{\mbox{\rm Im}}
\newcommand{\Ree}{\mbox{\rm Re}}
\newcommand{\Tr}{\mbox{\rm Tr}}
\newcommand{\Sp}{\mbox{\rm Sp}}
\newcommand{\Li}{\mbox{\rm Li}}
\newcommand{\MeV}{\mbox{\rm MeV}}
\newcommand{\GeV}{\mbox{\rm GeV}}
\newcommand{\fm}{\mbox{\rm fm}}

\newcommand{\with}{\mbox{\rm with}}
\newcommand{\while}{\mbox{\rm while}}
\newcommand{\annd}{\mbox{\rm and}}
\newcommand{\foor}{\mbox{\rm for}}
\newcommand{\oll}{\mbox{\rm all}}
\newcommand{\att}{\mbox{\rm at}}
\newcommand{\are}{\mbox{\rm are}}
\newcommand{\hc}{\mbox{\rm h.c.}}
\newcommand{\too}{\mbox{\rm to}}

\newcommand{\alphaQ}{\alpha_{\mbox{\rm eff}}(Q^2)}
\newcommand{\alphak}{\alpha_{\mbox{\rm eff}}(k_{E}^2)}
\newcommand{\alpham}{\alpha(\mu^2)}
\newcommand{\alphaq}{\alpha(Q^2)}
\newcommand{\al}{\alpha}
\newcommand{\als}{\alpha_{\mbox{\rm {\scriptsize s}}}}
\newcommand{\gs}{g_{\mbox{\rm {\scriptsize s}}}}
\newcommand{\muhad}{\mu_{\mbox{\rm {\scriptsize had.}}}}
\newcommand{\GF}{G_{\mbox{\rm {\tiny F}}}}
\newcommand{\MHA}{\mbox{\rm {\tiny MHA}}}

\newcommand{\qs}{\not \! q}
\newcommand{\xis}{\not \! \xi}
\newcommand{\gL}{\frac{1-\gamma_{5}}{2}}
\newcommand{\gR}{\frac{1+\gamma_{5}}{2}}
\newcommand{\gmut}{\mbox{$\tilde{\gamma}_{\mu}$}}
\newcommand{\smunut}{\mbox{$\tilde{\sigma}_{\mu\nu}$}}

\newcommand{\eff}{\mbox{\rm eff}}
\newcommand{\exxp}{\mbox{\rm exp.}}
\newcommand{\UV}{\mbox{\rm {\small UV}}}
\newcommand{\EM}{\mbox{\rm {\small EM}}}
\newcommand{\QCD}{\mbox{\rm {\footnotesize QCD}}}
\newcommand{\Lac}{\Lambda_{\chi}}
\newcommand{\msb}{\overline{\mbox{\rm\footnotesize MS}}}
\newcommand{\Lmsb}{\Lambda_{\overline{\mbox{\rm\footnotesize MS}}}}
\newcommand{\E}{\mbox{\rm {\tiny E}}}

\newcommand{\g}{\mbox{\bf g}}
\newcommand{\h}{\mbox{\bf h}}

\newcommand{\bP}{{\bf P}}
\newcommand{\bX}{{\bf X}}
\newcommand{\bQ}{{\bf Q}}
\newcommand{\Psls}{\not \! \bP}
\newcommand{\Qsls}{\not \! \bQ}
\newcommand{\ksls}{\not \! k}
\newcommand{\psls}{\not \! p}
\newcommand{\qsls}{\not \! q}
\newcommand{\pslsm}{\not \! p_{-}}
\newcommand{\pslsp}{\not \! p_{+}}

\newcommand{\stern}{\langle\bar{\psi}\psi\rangle}
\newcommand{\gew}{\mbox{\bf g}_{\mbox{\rm\footnotesize \underline{ew}}}}

\input epsf

\begin{titlepage}

\leftline{OUTP-09-11-P}

\vskip 1.5 cm

\begin{center}

{\Large On the possibility of light string resonances at the LHC\\}
{\Large and Tevatron from Randall-Sundrum throats\\}\vskip .3cm

\vskip 1.cm

{\large {Babiker Hassanain$^{a,b,\dag}$,
John March-Russell$^{a,\ddag}$, and
J. G. Rosa$^{a,\P}$}} 
\vskip 0.6cm

{\it $^a$ The Rudolf Peierls Centre for Theoretical Physics, \\
Department of Physics, University of Oxford. \\ 1 Keble Road,
Oxford, OX1 3NP, UK.} \\
\vskip 0.5cm
{\it $^b$ Christ Church College, Oxford, OX1 1DP, UK.} \\
\vskip 0.5 cm
$^\dag$ babiker@thphys.ox.ac.uk \\
$^\ddag$ jmr@thphys.ox.ac.uk    \\
$^\P$    rosa@thphys.ox.ac.uk \\
\vskip 0.5cm

\vspace{1.7cm}

\begin{abstract}
In string realizations of the Randall-Sundrum scenario, the higher-spin Regge excitations of
Standard Model states localized near the IR brane are warped down to close to the TeV scale. We argue that, as a consequence of the localization properties of Randall-Sundrum models of flavour, the lightest such resonance is the spin-3/2 excitation, $t_R^*$, of the right-handed top quark over a significant region of parameter space. A mild accidental cancellation allows the $t^*_R$ to be as light or lighter than the Kaluza-Klein excitations of the Standard Model states.  We consider from a bottom-up effective theory point of view the production and possible observability of such a spin-3/2 excitation at the LHC and Tevatron.  Current limits are weaker than might be expected because of the excess of $WWjj$ events at the Tevatron reported by CDF at $M_{\rm inv} \simeq 400-500$ GeV.   
\end{abstract}

\end{center}

\noindent

\end{titlepage}

\newpage






\section{Introduction}

Models with warped extra dimensions have been extensively studied since the seminal papers by Randall and Sundrum (RS) \cite{Randall:1999ee}.  Such models provide an elegant explanation of the 
hierarchy between the electroweak and Planck scales due to the exponential warping of the bulk geometry, which red-shifts scales of IR-localized physics down to the TeV scale.  Although in the original RS model only gravity propagates in the bulk, it was soon realized that Standard Model (SM) fields living in the warped extra dimension could provide new insights on other unsolved issues, such as the observed hierarchy of fermion masses. The latter arises in such models as a consequence of localizing fermionic wave-functions at different places in the fifth dimension, as discussed in \cite{Gherghetta:2000qt}.

The idea of warped extra dimensions also garnered much attention within superstring theory, as compactifications of ten-dimensional supergravity involving D-brane charges generically include regions with large warping, which became known as ``throats" \cite{Giddings:2001yu}, and warped throats are generic in the string theory landscape \cite{Dimopoulos:2001ui,Dimopoulos:2001qd,Cascales:2005rj,Giddings:2005ff,Hebecker:2006bn}. Although such solutions typically have a more complicated structure than the idealized RS model, many of the 5D phenomenological features can be reproduced in such string compactifications.  A particularly interesting construction embedding a probe D7 brane in the Klebanov-Witten throat, with an $AdS_5\times T^{1,1}$ near horizon geometry, includes bulk matter and gauge fields from strings ending on the probe brane \cite{Gherghetta:2006yq}. Although a non-trivial configuration with multiple branes is needed to implement the SM gauge group and chiral matter, this construction illustrates how string compactifications have the necessary ingredients to provide a top-down description of the RS bulk phenomenology (see also \cite{Acharya:2006mx}). Conversely, studies conducted within the RS model 
may be viewed as useful guides to the behaviour of warped supergravity compactifications. 

The string theory nature of warped throats suggests, however, that a much richer phenomenology should be included in these models, namely the effects of {\it higher-spin string excitations of SM fields}. In the context of large volume compactifications with intersecting brane configurations, such states have been shown to be sufficiently light to induce observable effects in current TeV-scale experiments \cite{Cullen:2000ef,Anchordoqui:2007da, Lust:2008qc, Anchordoqui:2008di, Anchordoqui:2009mm}.  In RS-like compactifications, where the bulk exponential warping generically brings Planckian mass states down to the TeV scale, one also naively
expects higher-spin string states to be brought down to near the TeV scale.  We argue that, although in warped models the generic string state is in fact expected to be somewhat heavier than the Kaluza-Klein (KK) mass of SM excitations, there can be individual higher-spin modes which are comparable in mass or even possibly lighter than the light KK excitations.  Given present experimental constraints on the KK mass-scale $M_{KK}$, we therefore do not expect the LHC to explore the kinematic regime $\sqrt{s} \gg M_{str}$ where soft Veneziano-like behaviour of cross-sections applies, but a regime where at most one or two Regge excitations might be accessible.

In this work, we focus on the possible observability of spin-3/2 excitations of SM fermions which correspond to the lightest higher-spin excitations of these particles in string theory.  In the absence of concrete string theory constructions, we follow a bottom-up approach to include these states in the RS scenario where fermions are allowed to propagate in the bulk.  Our analysis shows, under the assumption that string theory effects contribute universally in a flavour-blind fashion, that right-handed (rhd) top quark excitations ($t_R^*$) are the lightest of all fermion spin-3/2 resonances over a large region of parameter space.  This is a distinctive property of the RS scenario, unlike the universal spectrum that might be expected, eg, in large volume compactifications.  Not only is this feature easy to accommodate within present experimental data, which allows light top excitations with masses $\gsim 340$  GeV but puts stronger constraints on other lepton and quark excitations, but it also indicates a particular flavour signature in collider experiments.  Our results suggest that string spin-3/2 excitations of the top quark might be accessible at the LHC or Tevatron.  In fact current direct search limits on $t_R^*$ states are somewhat weaker than might be expected because of the excess of $WWjj$ events at the Tevatron reported by CDF at $M_{\rm inv} \simeq 400-500$ GeV \cite{Lister:2008is}. Interestingly, because of its enhanced production cross-section a spin-3/2 top resonance appears to be a better fit for this excess than a standard spin-1/2 top-prime. Such a spin-3/2 excitation might also be relevant to the excess reported by the D$\O$ collaboration \cite{D0note}.  

Turning to the outline of this paper, the spectrum of spin-3/2 resonances in the RS geometry is discussed in Section 2.  In Section 3,  we present the effective Lagrangian describing the interactions of $t_R^*$ with SM fermions, gauge fields and the Higgs field.  A preliminary discussion of the experimental signatures of singlet top quark excitations, particularly the production and detection prospects at the Tevatron and LHC follows in Section 4.  (We are also preparing a detailed analysis of such signals, the associated experimental background and possible methods for spin determination, to appear in a later publication \cite{future}.)  We conclude in Section 5.

\section{Spin-3/2 excitations in the Randall-Sundrum model}
\label{5Dspectrum}

With present techniques it is not possible to quantize strings on general curved backgrounds.  A famous exception
is IIB string theory on $AdS_5 \times S^5$, which by Maldacena duality is equivalent to maximally supersymmetric 
$SU(N)$ Yang-Mills theory in four dimensions \cite{Aharony:1999ti}.  In this case the string scale $M_{str}$ is parametrically heavier than the Kaluza-Klein scale $k$ 
arising from the $S^5$ by a factor $ M_{str}/k \sim (g_{str} N)^{1/4}$, and thus decouple from the theory in the formal $N$ and $g_{str} N \rightarrow \infty$ limits 
in which supergravity is a good description of the gravitational side of the duality.  However there are reasons to expect the situation to be different 
in the string realization of the RS scenario as a warped throat.  First, in RS model building, the curvature length $1/k$ of the $AdS_5$ space is not
much longer than the 5D Planck length, $M_5/k \sim N^{2/3} \sim {\rm few}$ being typical.  Since we 
expect (or require) the string coupling $g_{str} < 1$, we deduce that the string oscillators of a string realization of RS may be only slightly
heavier than the bulk KK modes. Second, the expected string realization of the RS scenario is not a throat
with the geometry of a perfect slice of $AdS_5$, but rather a so-called warped-defomed (or Klebanov-Tseytlin-like \cite{Klebanov:2000nc})
throat where the effective curvature length changes with distance $y$ along the throat, reducing in the
IR \cite{Brummer:2005sh,warpEWPT}.  We do not yet have a good understanding of the value of $M_5/k(y)$
in string realizations of realistic RS throats, so at present we take $M_5/k(y_{IR})$ to be a parameter of the
theory.  In fact $M_5/k(y_{IR}) \rightarrow {\cal O}(1)$ has significant potential phenomenological advantage in that the problems of the electroweak phase transition of RS models \cite{rattazzi} can be ameliorated \cite{warpEWPT}, and the $N$-enhanced contributions to precision electroweak observables possibly reduced to acceptable levels.   (Other papers which consider some of the phenomenological consequences of a Klebanov-Tseytlin-like throat with a smooth tip are \cite{Shiu:2007tn,McGuirk:2007er,Marchesano:2008rg}.) Note also that in a realistic realization of the Standard Model in a 5D throat (for example from intersecting D7-branes) the mass parameter in the effective field theory description of low-lying string states must depend on the details of the compactification leading to the 5D field theory. This may further reduce the 5D bulk mass of string resonances with respect to the 10D string scale $M_{str}$, although this is clearly a model-dependent statement. We believe that the end result is that the masses of low-lying string excitations are not necessarily much heavier than the KK modes at the IR tip of the throat. We now argue that flavour physics model building
in RS models in fact favours a light (compared to the other spin-3/2 Regge excitations) spin-3/2 resonance of the rhd top quark over much of parameter space, and in fact that this resonance can even be lighter than the KK modes if a mild
accidental cancellation occurs.  Since the flavour physics model building done so far has been performed exclusively
in a slice of $AdS_5$ we start by considering the mass spectrum of spin-3/2 Regge excitations of SM fermions living in the bulk of a RS five-dimensional space. 

The RS model \cite{Randall:1999ee} consists of a five-dimensional theory where the bulk geometry is a finite slice of AdS$_5$ space of length $L=\pi R$, truncated by two four-dimensional boundaries (branes) at $y=0$ and $y=L$, with $0\leq y\leq L$ representing the spatial coordinate along the finite extra-dimension\footnote{While originally this truncation corresponded to the five-dimensional orbifold $S^1/\mathbb{Z}_2$ with four-dimensional branes at the orbifold fixed points, this formulation is more suitable for our discussion.}. The metric is then given by:
\begin{equation} \label{RS metric}
ds^2=e^{-2ky}\eta_{\mu\nu}dx^{\mu}dx^{\nu}+dy^2~,
\end{equation}
where $\eta_{\mu\nu}$, $\mu=1,\dots,4$, is the 4-dimensional Minkowski metric with mostly plus signature and $k$ is the curvature of the anti-de Sitter space. The exponential ``warp" factor then induces a large hierarchy between the effective mass scales at the two boundaries, $M_P$ and $M_Pe^{-kL}$. If the former is the fundamental Planck scale, one obtains an $\cal{O}$(TeV) mass scale at $y=L$ for $kR\sim 12$. The RS model thus explains the large hierarchy between gravitational and electroweak interactions if the Higgs field is localized in the ``TeV-brane".

Let us then consider a massive five-dimensional Dirac vector-spinor field $\Psi_{M}$, $M=0,\ldots,3,5$, described by the 5D extension of the Rarita-Schwinger action \cite{Kogan:2001wp}\footnote{Our definition of $M_\Psi$ differs from that in \cite{Kogan:2001wp}, corresponding in our case to the physical pole in the 5D propagator.}:
\begin{eqnarray} \label{Rarita-Schwinger5D}
S_5=\int d^4x\int dy\sqrt{-G}\bar{\Psi}_M\Gamma^{MNP}\bigg(D_N+{1\over3}M_{\Psi}\Gamma_N\bigg)\Psi_P~,
\end{eqnarray}
where the covariant derivative is given by
\begin{eqnarray}
D_M\Psi_N=\partial_M\Psi_N-\Gamma^P_{MN}\Psi_P+{1\over2}\omega^{AB}_M\gamma_{AB}\Psi_N~,
\end{eqnarray}
with $\Gamma^P_{MN}$ denoting the five-dimensional Christoffel symbols and $\omega^{AB}_M$ the components of the associated spin-connection. In Eq.(\ref{Rarita-Schwinger5D}), the 5D curved space gamma matrices are related to the corresponding flat space representation $\gamma^a$ via $\Gamma_M=e_M^a\gamma_a$, where $e^a_\mu=e^{-ky}\delta^a_{\mu}$ for $a,\mu=0,\ldots,3$, $e^5_5=1$ and all other components of the vielbein vanish. We have also defined the antisymmetric product $\Gamma^{MNP}=\Gamma^{[M}\Gamma^N\Gamma^{P]}$. We assume the mass term is generated in a gauge invariant way with respect to the spinorial transformation $\Psi_M\rightarrow \Psi_M+\partial_M\epsilon$ where $\epsilon$ is a spinorial gauge parameter, and parametrize the bulk mass as $M_\Psi=ck$ for some real constant $c$. We may also use this gauge invariance to set $\Psi_5=0$ \cite{Kogan:2001wp}, so that it suffices to denote the field by $\Psi_{\mu}$, $\mu=0,\dots,3$.

As for the case of bulk spin-1/2 fermions, it is convenient to separate the vector-spinor in its chiral components $\Psi_{\mu}^{L,R}=\pm\gamma_5\Psi_{\mu}^{L,R}$. These admit a KK decomposition of the form:
\begin{eqnarray} \label{KK decomposition}
\Psi_{\mu}^{L,R}(x,y)={1\over \sqrt L}\sum_n\psi_{\mu n}^{L,R}(x)e^{2ky}f_n^{L,R}(y)~,
\end{eqnarray}
where the KK mode-functions are normalized as:
\begin{eqnarray} \label{KK normalization}
{1\over L}\int_0^Ldye^{ky}f_m^{L,R*}f_n^{L,R}=\delta_{mn}~.
\end{eqnarray}

Using the equations of motion derived from the Rarita-Schwinger action, one concludes that the right- and left-handed mode functions satisfy the coupled differential equations:
\begin{eqnarray} \label{mode equations}
(\partial_5+M_{\Psi})f_n^R&=&m_ne^{ky}f_n^{L}~,\nonumber\\
(-\partial_5+M_{\Psi})f_n^L&=&m_ne^{ky}f_n^{R}~,
\end{eqnarray}
where $m_n$ is the 4D (Dirac) mass of the $n-$th KK mode. These are in fact the same equations satisfied by the KK mode functions of a bulk spin-1/2 field \cite{Gherghetta:2000qt}, which is not surprising since the Rarita-Schwinger vector-spinor satisfies the five-dimensional Dirac equation. It is convenient to define $t=e^{k(y-L)}$, so that $\epsilon\leq t\leq1$, where $\epsilon=e^{-kL}\sim10^{-16}$ in order to solve the hierarchy problem. In terms of this variable, the KK mode equations can be written as:
\begin{eqnarray} \label{t mode equations}
t^2\partial_t^2f_n^{L,R}+\big(x_n^2t^2-c(c\mp1)\big)f_n^{L,R}=0~,
\end{eqnarray}
where $x_n\equiv(m_n/k\epsilon)\sim(m_n/\mathrm{TeV})$ and the + (--) sign corresponds to the right-handed (left-handed) modes. These equations have the general solution:
\begin{eqnarray}
f_n^{R}(t)&=&\sqrt{t}\big[a_n^RJ_{c+{1\over2}}(x_nt)+b_n^RJ_{-c-{1\over2}}(x_nt)\big]~,\nonumber\\
f_n^{L}(t)&=&\sqrt{t}\big[a_n^LJ_{{1\over2}-c}(x_nt)+b_n^LJ_{c-{1\over2}}(x_nt)\big]~,
\end{eqnarray}
where $J_{\alpha}(x)$ represents a Bessel function of the first kind. Eq.(\ref{mode equations}) then implies $a_n^R=b_n^L$ and $a_n^L=-b_n^R$. 

To determine the KK spectrum, one needs to impose boundary conditions (b.c.) for the mode functions in the UV ($t=\epsilon$) and IR ($t=1$) branes. Bulk quarks and leptons, which have analogous KK mode solutions, satisfy the same b.c. in both branes (either Dirichlet or Neumann), with opposite b.c. for left- and right-handed modes, according to Eq.(\ref{mode equations}).  If this were also the case for their spin-3/2 excitations, it would necessarily lead to 4D massless chiral vector-spinors. Then, through Yukawa couplings to the Higgs field, these would typically acquire masses of the same order of their SM spin-1/2 counterparts, which is phenomenologically unacceptable.   However, one is free to impose distinct b.c. in the two branes. This eliminates all spin-3/2 chiral zero-modes, the lightest resonances corresponding to the first KK modes of the underlying five-dimensional fields.  

We may then have two possible b.c. assignments for left-handed modes, which we denote by (a) (--,+) and (b) (+,--), where -- refers to Dirichlet and + to Neumann b.c. in the (UV,IR) branes.  As $\epsilon\ll1$, the KK masses are approximately given by the zeros of Bessel functions:
\begin{eqnarray} \label{KK masses}
(a)\begin{cases}
J_{{1\over2}+c}(x_n)\simeq0, &c\geq{1\over2}\\
J_{-{1\over2}-c}(x_n)\simeq0, &c<{1\over2}
\end{cases}\qquad~,\qquad
(b)\begin{cases}
J_{-{1\over2}+c}(x_n)\simeq0, &c\geq-{1\over2}\\
J_{{1\over2}-c}(x_n)\simeq0, & c<-{1\over2}
\end{cases}~.
\end{eqnarray}

\begin{figure}[htbp] 
	\centering
		\includegraphics[scale=0.8]{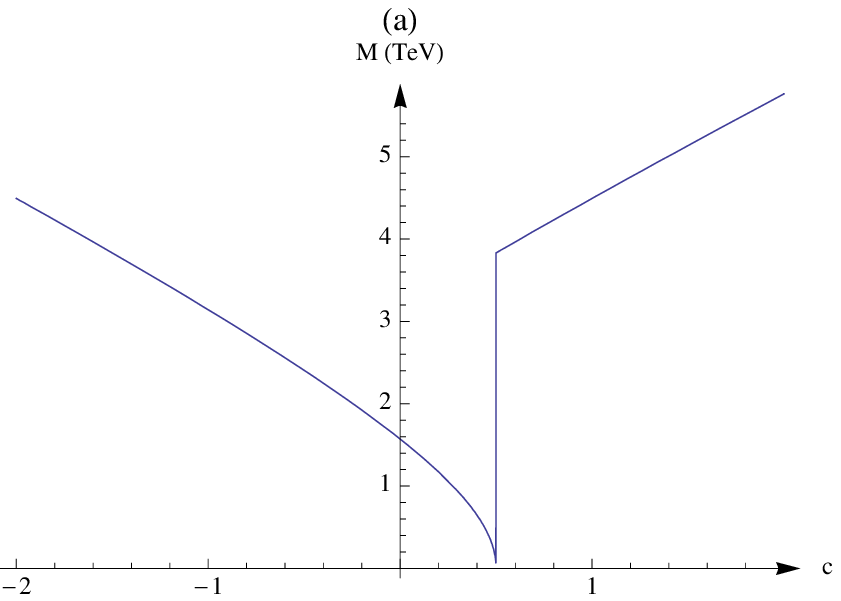}$\qquad\qquad$
		\includegraphics[scale=0.8]{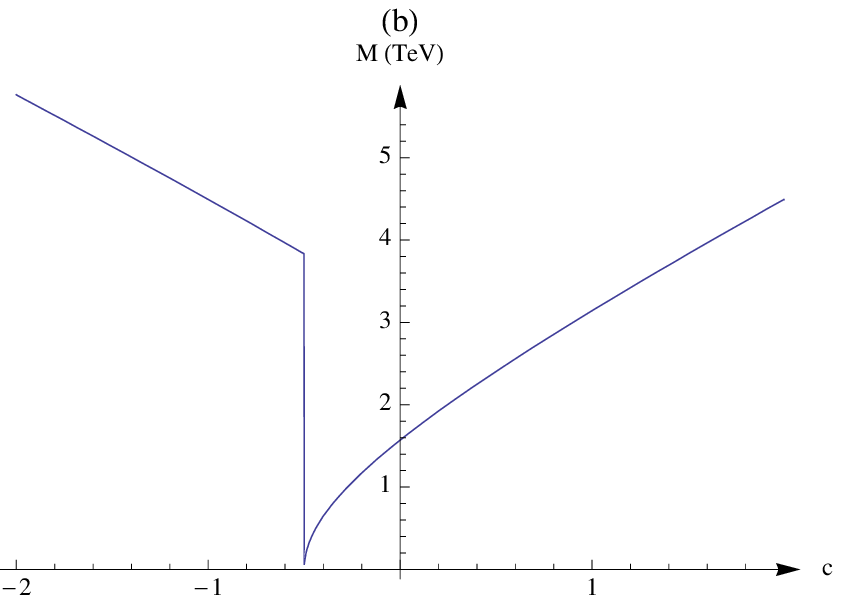}
		\caption{Mass of the first KK mode of a spin-3/2 resonance for (a) (--,+) and (b) (+,--) mixed boundary conditions as a function of the bulk mass parameter $c$.}
		\label{KK masses generic}
\end{figure}

Note that the two types of b.c. are related by the redefinition $c\rightarrow -c$. In Figure \ref{KK masses generic} we plot the values of the first KK mode mass as a function of the bulk mass parameter $c$. As one may observe, there exist values of $c$ for which this mode can become quite light.

We have, however, made no assumptions about the values of the bulk mass parameter so far. Quark and lepton bulk masses must be distinct to explain the observed 4D fermion mass hierarchy \cite{Gherghetta:2000qt}. In string theory, these states are massless in the decompactification limit and have an associated tower of higher-spin states with a mass gap of the order of the fundamental string scale. Compactification into a five-dimensional RS throat should, hence, induce bulk masses for the spin-1/2 ground states, subsequently affecting all higher-spin excitations. There is, however, no fundamental reason to assume the mass gap between the resonances is affected by the mechanism that produces bulk fermion masses. Thus, we expect the bulk mass of the spin-3/2 resonances to include the mass of its spin-1/2 counterpart, $c_{0i}k$, and an extra string theory contribution, $sk$, which we assume to be the same for all spin-3/2 resonances and, following the discussion at the start of this section, also of the order of the fundamental 5D mass scale, so that $c_i=c_{0i}+s$.

The bulk masses of SM quarks and leptons determine their Yukawa couplings and hence their masses and CKM mixing, and phenomenology requires these masses to be close to the values $-k/2$ and $k/2$ for $SU(2)_W$ doublet and singlet fields, respectively \cite{Casagrande:2008hr}. For a single generation of quarks $q,u,d$ it is then useful to define the bulk mass parameters as $c_{0q}\equiv M_q/k$ and $c_{0u,d}\equiv-M_{u,d}/k$. The corresponding 4-dimensional Yukawa coupling is then related to the respective 5-dimensional coupling by:
\begin{eqnarray} \label{Yukawa 5D}
\lambda_{u,d}=\lambda_5k\bigg({1/2+c_{0u,d}\over e^{(1+2c_{0u,d})kL}-1}\bigg)^{1/2}\bigg({1/2+c_{0q}\over e^{(1+2c_{0q})kL}-1}\bigg)^{1/2}e^{(1+c_{0u,d}+c_{0q})kL}~,
\end{eqnarray}
where one assumes $\lambda_5k\sim1$. Note that this result neglects the effects of brane-localized operators in the 5-dimensional equations of motion, which is a good approximation for the fermion mass spectrum \cite{Casagrande:2008hr}. Eq.(\ref{Yukawa 5D}) illustrates how UV-localized fermions, with  $c_{0q,u,d}\lesssim -1/2$, lead to exponentially small Yukawa couplings, hence describing the masses of first and second generation quarks and also all lepton generations. The large top Yukawa coupling requires significantly larger bulk mass parameters, in particular for the singlet component. In \cite{Gherghetta:2000qt}, it was shown that UV localization of light fermions suppresses their couplings to KK gauge bosons, making them universal and hence preventing large FCNC effects. The latter might then be significant only for third generation quarks, but constraints from flavour-changing processes are much weaker in that case.

\begin{figure}[htpb] 
\begin{center}
		\includegraphics[scale=0.97]{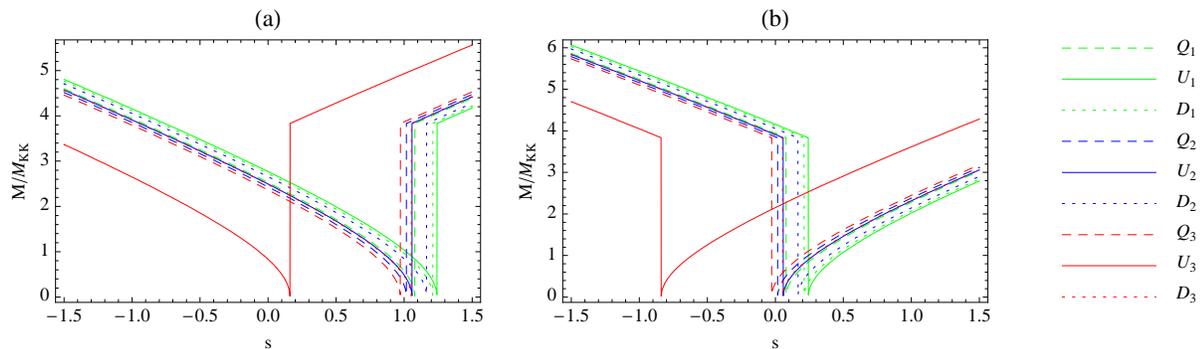}
		\caption{Mass of the first KK mode of spin-3/2  quark excitations as a function of the string bulk mass contribution for (a) (--,+) and (b) (+,--) b.c. for left-handed components of doublet modes, with singlet modes satisfying opposite boundary conditions. Masses are normalized to the nominal KK scale, $M_{KK}=k\epsilon\sim \mathcal{O}(\mathrm{TeV})$.}
		\label{quark_resonances}
		\end{center}
\end{figure}

Experimental data cannot, however, fully determine all SM bulk mass parameters, preventing a precise computation of the spin-3/2 resonance spectrum. One expects, however, all allowed spectra to exhibit a large separation between the mass of singlet top excitations and all remaining spin-3/2 resonances, due to the required difference between their bulk masses. To illustrate this point, we have used the set of quark bulk mass parameters given in \cite{Casagrande:2008hr} (see also \cite{Huber:2003tu} and \cite{Agashe:2004cp}), which is consistent with all experimental constraints concerning quark masses and mixing, to compute the first KK masses of the associated spin-3/2 resonances, $Q_i,\ U_i$ and $D_i$, $i=1,2,3$. Our results are shown in Figure \ref{quark_resonances}.

Recall that, in Eq.(\ref{Yukawa 5D}), we defined the bulk mass parameters for singlet and doublet excitations with opposite signs. Consequently, each plot in Figure \ref{quark_resonances} depicts singlet and doublet fields which satisfy opposite b.c., with plot (a) corresponding to the same IR b.c. satisfied by all spin-1/2 fermions.

Although not shown explicitly, lepton resonances will generically lie close to the first and second generation quark excitations, as they have similar bulk mass parameters. These results clearly show that the singlet top excitation is significantly lighter (heavier) than all others for negative (positive) $s$. The sign of the string bulk mass parameter is undetermined in the absence of a concrete string compactification, although negative values may naively seem more natural given that $c_0\lesssim -1/2$ for most fermions. Also, there is a sizeable range of $s$ for which the singlet top excitation is lighter than the KK mass scale (the lightest KK excitations have mass $\simeq 2.45 M_{KK}$ in RS models). As we will discuss in the next sections, it is phenomenologically more interesting and viable to have light singlet top excitations while all other resonances are much heavier, so that we will focus on this possibility for the remainder of this work.

One must also take into account that brane-localized Yukawa operators may modify the spectrum of spin-3/2 resonances after electroweak symmetry breaking. Such operators mix doublet and singlet states via the Higgs field and may (i) involve only spin-3/2 fields or (ii) mix spin-1/2 and spin-3/2 fields. However, their effects are naturally suppressed by powers of $v/M_{KK}$, where $v$ denotes the Higgs vacuum expectation value, and may thus be neglected to lowest order. Furthermore, operators of type (i) vanish to leading order when doublet and singlet states satisfy the same b.c. in the IR. The same happens for operators of type (ii) in the case where spin-3/2 resonances have the same IR b.c. as their spin-1/2 counterparts, which could be interpreted as a natural consequence of the intrinsic UV nature of string theory effects.

\section{Spin-3/2 rhd top interactions}
\label{resonance-Interactions}

Motivated by the spectrum of spin-3/2 resonance states outlined in Section~2 we will focus on the case where the sole light ($M < {\rm few~TeV}$) spin-3/2 state is the partner of the rhd top, which we denote by $t_R^*$.  
Such states are least constrained by current direct searches, but as we will show below can still be copiously
produced at the LHC, and have the added advantage of relatively easily evading flavour-changing and rare-decay constraints.

Although the form of the low energy interactions of spin-3/2 resonances should ultimately arise from the underlying theory at high energies, such as string theory or a confining gauge theory leading to composite quark and leptons, we follow a bottom-up effective-field-theory approach and use a generic interaction Lagrangian involving the resonances and respecting the gauge symmetries of the SM.  Such an effective field theory must break down at high energy due to the non-renormalizable nature of interacting spin-3/2 theories.  Thus our theory is endowed with a cutoff $\Lambda$ (the maximum value of which we will later estimate), which suppresses higher-dimension operators\footnote{Theories of interacting spin-3/2 fields have been studied in the literature, mainly in the context of processes involving the $\Delta$ baryon-resonances \cite{Peccei:1968,Nath:1979wr,Benmerrouche:1989uc,Luty:1993fu,Kambor:1996xh,Pascalutsa:1998pw,Pascalutsa:1999zz,Napsuciale:2006wr}, or as gravitinos in supergravity \cite{Ferrara:1976fu}, although studies of interacting spin-3/2 resonances of SM quarks and leptons can also be found \cite{Walsh:1999pb,Kuhn:1984rj,Moussallam:1989nm,Dicus:1998yc,Almeida:1995yp,Burges:1983zg,ArkaniHamed:1997jv,Babu:1997jr}.}.

We start by considering a generic Dirac vector-spinor $\psi_\mu$ of mass $M$ in four dimensions, transforming in the fundamental representation of a local gauge group $G$ and representing the first KK mode of an underlying five-dimensional spin-3/2 field. A minimally coupled vector-spinor is described by the gauge-invariant Rarita-Schwinger Lagrangian \cite{Rarita:1941mf}, defined up to a parameter $A$ which is related to the presence of unphysical spin-1/2 degrees of freedom in the vector-spinor field $\psi_\mu$ (see eg \cite{Napsuciale:2006wr}). We take the value $A=-1$, so that:
\begin{eqnarray} \label{minimal}
\mathcal{L}_4=i\bar{\psi}_{\mu}\gamma^{\mu\nu\rho}D_{\nu}\psi_{\rho}+M\bar{\psi}_\mu\gamma^{\mu\rho}\psi_{\rho}~,
\end{eqnarray}
with the standard gauge covariant derivative $D_{\mu}=\partial_{\mu}-igA_{\mu}$, where $g$ is the gauge coupling and $A_{\mu}$ the gauge field.   Following from this $\psi_{\rho}$ satisfies 
the Dirac equation.  The free Lagrangian also implies the constraint
\begin{equation} \label{constraint}
\gamma^\rho \psi_{\rho} = 0 \, ,
\end{equation}
which, however, is violated when interactions with $A_\mu$ are included.
The free Lagrangian yields the propagator \cite{Moussallam:1989nm}:
\begin{equation} \label{propagator}
\mathcal{P}^{\mu\nu}=\frac{1}{p^2-M^2}\left[-(\not{p}+M)\left(\eta^{\mu\nu}-\frac{p^\mu p^\nu}{M^2}\right)
-\frac{1}{3}\left(\gamma^\mu+\frac{p^\mu}{M}\right)(\not{p}-M)\left(\gamma^\mu+\frac{p^\nu}{M}\right)\right] \, ,
\end{equation}
where the expression in square brackets gives the associated polarization tensor. As excited fermions, the spin-3/2 resonances have the same SM quantum numbers and gauge couplings as their spin-1/2 counterparts which, in the
case of $t_R^*$, correspond to the $(3,1)_{2/3}$ representation of the SM gauge group.

The numerator of the propagator Eq.(\ref{propagator}) contain terms that go like three powers of the momentum.  This leads to cross-sections which go like positive powers of the energy in the ultra-violet, an observation which is symptomatic of the field theory of  higher-spin particles. As the cross-section rises indefinitely, tree-level unitarity is violated at some point.  One can get a rough estimate of the unitarity bound by simply considering the amplitude of elastic scattering in the s-wave only.  For gluon-mediated interactions arising from Eq.(\ref{minimal}), one finds that tree-level unitarity is respected for $\sqrt{s} \lsim 7 M$.  Of course it is possible for the effective theory describing the resonance to break down before this scale, but as a working hypothesis we will take $\Lambda = 7 M$ unless otherwise specified.   

As we will see, Eq.(\ref{minimal}) already contains the leading order interaction responsible for the dominant production mode of the spin-3/2 resonances at the Tevatron and the LHC, but does not allow for its decay. Therefore we are also interested in interactions involving both spin-3/2 and spin-1/2 states, which allow the decay of the resonances into SM fermions.  These will {\it a priori} include dimension-four ``kinetic" and ``mass mixing" interactions of the form
\begin{eqnarray}\label{d4interactions}
\mathcal{L}_{4}^{\mathrm{mix}}&=&i\alpha_1\bar\psi_{\mu}(\eta^{\mu\nu}+w\gamma^{\mu}\gamma^{\nu})D_\nu P_{R,L}\chi+\alpha_2\bar\psi_{\mu}\gamma^\mu P_{R,L}\chi+\mathrm{h.c.}~,
\end{eqnarray}
where $P_{R,L}\chi$ denotes a chiral spin-1/2 state with the same quantum numbers as the $\psi_\mu$ excitation. As for KK excitations of SM fermions or generically any model of new vector-like quark and lepton families, one may always choose a basis where kinetic and mass terms are diagonal, so that terms such as Eq.(\ref{d4interactions}) are unphysical. However, we must also consider the Yukawa operators mentioned in the previous section, which are of the form 
\begin{equation} \label{Higgs}
\mathcal{L}_{Higgs}= \lambda~ \bar\chi  ~i\sigma_2 H^*  ~\gamma_\mu\psi^{\mu} + \mathrm{h.c.}~,
\end{equation}
where $\chi$ represents a lhd quark $SU(2)_W$ doublet in the case where $\psi_\mu$ represents the $t_R^*$ resonance. These operators are not necessarily diagonal in the above mentioned basis, so that generically there will be dimension-four mixing between spin-1/2 and spin-3/2 mass eigenstates in weak interactions. 

Although the interactions with the physical Higgs boson vanish for on-shell spin-3/2 states due to the constraint Eq.(\ref{constraint}), for spin-3/2 resonances appearing in internal lines the new Higgs-mediated interactions may in general lead to flavour violating rare processes.  However, these FCNC do not severely constrain the couplings $\lambda$ or the masses of the $t_R^*$ state, because their FCNC effects are suppressed by the same mechanism that suppresses the couplings of KK fermions to SM fermions and Higgs in Randall-Sundrum models of flavour physics \cite{Casagrande:2008hr, Agashe:2004cp, Blanke:2008zb}. Also, as discussed in the previous section, the effects of these operators are small for $v \ll M_{KK}$, where $v$ is the VEV of the Higgs, and may be further suppressed depending on the choice of b.c. in the IR. Hence, we expect all dimension-four weak operators mixing spin-1/2 and spin-3/2 states to be suppressed. 

This mixing will be absent in strong interactions involving spin-1/2 and spin-3/2 states, which may only occur via dimension-five operators of the form:
\begin{eqnarray}\label{chiral interactions}
\mathcal{L}_{5}&=&i{a\over\Lambda}\bar\psi_{\mu}(\eta^{\mu\alpha}+z\gamma^{\mu}\gamma^{\alpha})F_{\alpha\beta}\gamma^{\beta}P_{R,L}\chi+\mathrm{h.c.}~,
\end{eqnarray}
where for $SU(2)_W$ singlet excitations terms involving lhd SM fermions are further suppressed by $v/\Lambda$.  Here we assume that the mass scale $\Lambda$ sets the strength of the dimension-five operators, and $a$ is an ${\cal O}(1)$ coefficient that one may expect to scale with the associated coupling constant $a\sim g$.\footnote{The parameter $z$ is the so-called \textit{off-shell parameter}, as the constraint $\gamma^{\mu}\psi_{\mu}=0$ eliminates the contribution of this interaction for physical spin-3/2 states. Several discussions on the value of this parameter can be found in the literature (see eg \cite{Benmerrouche:1989uc}). A popular choice is the value $z=-{1\over4}$, first obtained in \cite{Peccei:1968}. Other authors have found from field-theoretical arguments the value $z=-{1\over2}$  \cite{Nath:1979wr}, while some phenomenological approaches use $z=0$ \cite{Kuhn:1984rj}. All of these various specifications of $z$ rely on effective field theories with several formal problems, and, given the observed discrepancy, we do not believe a solid value of $z$ can be established using this type of approach and leave it as a free parameter of the interaction Lagrangian.   Because of the physical constraint $\gamma^{\mu}\psi_{\mu}=0$ the value of $z$ affects neither the decay width of the resonance at leading order, nor the dominant (pair) production mechanism.}  For singlet excitations, the interaction Eq.(\ref{chiral interactions}) contains, in addition to the gluonic field strength, couplings to the photon and $Z$, which, if $a_i\sim g_i$, are sub-dominant. The strong interactions in Eq.(\ref{chiral interactions}) are not necessarily flavour-diagonal, but it is easy to show that, assuming that in the 5D AdS theory the suppression of higher dimensional operators is by  powers of the 5D Planck mass $M_5$, couplings of the light quarks to the resonances will be more strongly suppressed than those of the top. Another way to express this is to say that the light-quarks are UV-localized, and thus exposed to higher mass-scales than the IR-localized top quark. Thus, the coupling of $t_R^*$ to the top-quark will be dominant compared to other possible flavour-violating couplings. As we will see later, the gluonic operator in Eq.(\ref{chiral interactions}) gives the dominant contribution for single-production of the $t_R^*$ resonance at the LHC. 
 
We conclude this section by noting that one can attempt to constrain the masses and couplings of the spin-3/2 resonances via low-energy observables and precision tests. These will be directly analogous to the constraints applicable to the KK modes of the SM fermions in usual RS model-building \cite{Casagrande:2008hr}. However the spin-3/2 sector fields (with the exception of $t_R^*$) are generically heavier than the fermion KK modes, and their mixings with the SM fermions can be naturally suppressed, so that it is easy to accommodate the constraints for large regions of parameter space.  Specifically, the vector-like $SU(2)_W$ singlet $t_R^*$ is loosely constrained by low-energy observables and precision tests such as the oblique electroweak observables $S,T$ and $U$, four-fermion operators, FCNC observables like $\epsilon'/\epsilon$ \cite{Gedalia:2009ws} and rare decays like $b\rightarrow s\gamma$ \cite{Agashe:2008uz}, even if its mass is below the TeV-scale. Although such observables might help constraining the mass of this state, a detailed analysis cannot be done in a model-independent way, lying beyond the scope of this paper. Better constraints can, however, be obtained by direct searches, a topic we discuss in the next section.

\section{Experimental Signatures}
\lbl{decay}

We expect the dominant decay modes of $t_R^*$ to be into a rhd top and a gluon, via the interaction Eq.(\ref{chiral interactions}) (in accord with the results of string theory calculations in flat space \cite{Anchordoqui:2008hi}), in addition to  $t_R^*\to \gamma t$ via analogous dimension five interaction terms, and $t_R^* \to H t$, $t_R^* \to Z t$ and $t_R^*\to W^+d_i$ via the suppressed dimension four mixings Eqs.(\ref{d4interactions}) and (\ref{Higgs}) (see \cite{Barger:1995dd} and references therein for the related case of a spin-1/2 singlet excitation). Despite the dimension-five nature of the top+gluon channel, it corresponds to a strong interaction, while the last three decay channels occur via dimension-four interactions suppressed by small mixing angles for $v\ll M_{KK}$, and possibly further suppressed for specific choices of IR b.c., as discussed earlier. Resonances with smaller KK masses should be more sensitive to IR effects, so one expects a larger fraction of weak decays in this case, although in general the strong and weak channels may have comparable branching ratios. Note that the weak decay channels will get other contributions from transition-magnetic-moment-like interactions as given in Eq.(\ref{chiral interactions}). The various decay channels may lead to distinctive final states useful for distinguishing $t_R^*$ production from the SM background, but unfortunately their branching ratios depend on the unknown coefficients $a_i$ for $SU(3)$ and $U(1)_Y$ and on the Yukawa couplings. As the $t_R^* \rightarrow t g$ partial width is $\Gamma_{tg}\simeq a^2 M^3/(48 \pi\Lambda^2)$, where $M$ is the mass of the resonance, and given that we expect $M<\Lambda$, it is clear that $\Gamma_{tg} \ll M$. This should also hold for the weak decay channels, as for spin-1/2 vector-like excitations \cite{Atre:2008iu}. Hence, this allows us to utilize the narrow-width approximation (NWA), thus considering the production and decay of the resonance as a two-step process.
 
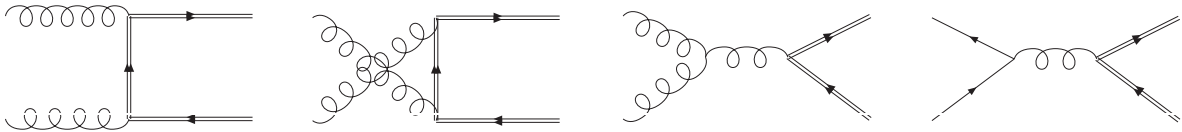
\begin{figure}[htbp]
\begin{center}
\fcolorbox{white}{white}{
  \begin{picture}(614,40) (13,-47)
    \SetWidth{0.5}
    \SetScale{0.72}
    \SetScaledOffset(40,-30)
    \SetColor{Black}
    \Gluon(13.47,11.67)(78.12,11.67){5.39}{5.14}
    \ArrowLine(78.12,12.57)(142.78,12.57)\ArrowLine(78.12,10.78)(142.78,10.78)
    \ArrowLine(142.78,-43.1)(78.12,-43.1)\ArrowLine(142.78,-41.31)(78.12,-41.31)
    \Gluon(13.47,-42.21)(78.12,-42.21){-5.39}{4.29}
    \ArrowLine(77.23,-42.21)(77.23,11.67)\ArrowLine(79.02,-42.21)(79.02,11.67)
    \ArrowLine(238.86,11.67)(303.52,11.67)\ArrowLine(238.86,9.88)(303.52,9.88)
    \Gluon(174.21,-43.1)(238.86,10.78){-5.39}{6.86}
    \Gluon(174.21,10.78)(238.86,-43.1){5.39}{6.86}
    \ArrowLine(303.52,-44)(238.86,-44)\ArrowLine(303.52,-42.21)(238.86,-42.21)
    \ArrowLine(237.97,-43.1)(237.97,10.78)\ArrowLine(239.76,-43.1)(239.76,10.78)
    \Gluon(379.85,-9.88)(422.95,-9.88){5.39}{2.57}
    \Gluon(336.74,11.67)(379.85,-9.88){5.39}{3.43}
    \Gluon(336.74,-42.21)(379.85,-9.88){-5.39}{3.43}
    \ArrowLine(422.55,-9.07)(465.65,12.48)\ArrowLine(423.35,-10.68)(466.46,10.87)
    \ArrowLine(465.52,-42.92)(422.41,-10.6)\ArrowLine(466.59,-41.49)(423.49,-9.16)
    \ArrowLine(541.48,-10.78)(498.38,10.78)
    \ArrowLine(498.38,-43.1)(541.48,-10.78)
    \ArrowLine(584.19,-9.97)(627.29,11.58)\ArrowLine(584.99,-11.58)(628.09,9.97)
    \ArrowLine(628.23,-42.38)(585.13,-10.06)\ArrowLine(627.15,-43.82)(584.05,-11.49)
    \Gluon(541.48,-10.78)(584.59,-10.78){5.39}{2.57}
  \end{picture}
}
\end{center}
\caption{Feynman diagrams contributing to pair-production of singlet top spin-3/2 resonances (double lines) via gluon-fusion and quark-antiquark annihilation (except $t\bar{t}$).}
\label{feyn_pair}
\end{figure}

Armed with the justification of the NWA, we may safely proceed to the calculation of the production cross-sections of the resonance in proton-proton (and proton-antiproton) collisions. As we will see, the dominant production mechanism is typically pair-production via the interaction of Eq.(\ref{minimal}).  The contributing diagrams 
coming from this operator are shown in Figure \ref{feyn_pair}. 

In principle, there are diagrams in which the dimension-five operator of Eq.(\ref{chiral interactions}) contributes, but these are strongly suppressed for the values of $\Lambda$ we use below. We note that pair-production of spin-3/2 states has been considered by \cite{Moussallam:1989nm,Dicus:1998yc}. We find, consistently with the results of those authors, that the cross-sections of pair-production at the LHC are quite large. For example, the production cross-section for a resonance mass of $300$ GeV is of the order of $10^3$ pb, while a minimally coupled spin-1/2 particle of the same mass would be pair-produced with cross-sections of the order of $10$ pb. This difference is due 
to the fact that the partonic cross-sections for the spin-3/2 particle go like $s^3$ in the ultraviolet, whereas those of a spin-1/2 particle go like $s^{-1}$. Therefore, upon convolution 
with the parton distribution functions (PDF) to obtain the hadronic cross-sections, the integral picks up more contributions from the large $x$ region, where $x$ is the momentum fraction available to the partons. This violent ultraviolet behaviour also means that pair-production violates unitarity at some scale, which we estimate 
to be $\sqrt{s} \approx 7 M$, where $M$ is the mass of the resonance. A simple way to parameterize the inevitable slop in this scale is to impose a hard cut-off $\sqrt{s_0}$ on the 
partonic cross-sections at a certain value of the centre-of-mass energy.  

\begin{figure}[htbp]
\begin{center}
\fcolorbox{white}{white}{
  \begin{picture}(614,100) (13,-46)
    \SetWidth{0.5}
    \SetScale{0.72}
    \SetScaledOffset(40,-30)
    \SetColor{Black}
    \Gluon(13.48,105.16)(78.19,105.16){5.39}{5.14}
    \Gluon(13.48,51.23)(78.19,51.23){-5.39}{4.29}
    \ArrowLine(77.29,51.23)(77.29,105.16)\ArrowLine(79.09,51.23)(79.09,105.16)
    \ArrowLine(78.19,106.06)(142.9,106.06)\ArrowLine(78.19,104.26)(142.9,104.26)
    \ArrowLine(142.9,51.23)(78.19,51.23)
    \Gluon(337.04,104.26)(401.75,104.26){5.39}{5.14}
    \Gluon(337.04,50.33)(401.75,50.33){-5.39}{4.29}
    \ArrowLine(401.75,50.33)(401.75,104.26)
    \ArrowLine(401.75,105.16)(466.46,105.16)\ArrowLine(401.75,103.36)(466.46,103.36)
    \ArrowLine(466.46,50.33)(401.75,50.33)
    \Gluon(498.82,104.26)(563.53,50.33){5.39}{6.86}
    \Gluon(498.82,50.33)(563.53,104.26){-5.39}{6.86}
    \ArrowLine(563.53,105.16)(628.24,105.16)\ArrowLine(563.53,103.36)(628.24,103.36)
    \ArrowLine(563.53,50.33)(563.53,104.26)
    \ArrowLine(628.24,50.33)(563.53,50.33)
    \Gluon(56.62,-12.58)(99.76,-12.58){5.39}{2.57}
    \Gluon(13.48,8.99)(56.62,-12.58){5.39}{3.43}
    \Gluon(13.48,-44.94)(56.62,-12.58){-5.39}{3.43}
    \ArrowLine(99.36,-11.78)(142.5,9.79)\ArrowLine(100.17,-13.39)(143.31,8.18)
    \ArrowLine(142.9,-44.94)(99.76,-12.58)
    \Gluon(175.26,9.89)(239.97,-11.68){5.39}{5.14}
    \Gluon(175.26,-44.04)(239.97,-11.68){-5.39}{5.14}
    \ArrowLine(239.69,-10.83)(304.4,10.74)\ArrowLine(240.26,-12.54)(304.97,9.03)
    \ArrowLine(304.68,-44.04)(239.97,-11.68)
    \ArrowLine(380.18,-13.48)(337.04,8.09)
    \ArrowLine(337.04,-45.84)(380.18,-13.48)
    \ArrowLine(422.92,-12.68)(466.06,8.89)\ArrowLine(423.72,-14.29)(466.86,7.29)
    \ArrowLine(466.46,-45.84)(423.32,-13.48)
    \Gluon(380.18,-13.48)(423.32,-13.48){5.39}{2.57}
    \ArrowLine(239.07,105.16)(303.78,105.16)\ArrowLine(239.07,103.36)(303.78,103.36)
    \Gluon(174.36,50.33)(239.07,104.26){-5.39}{6.86}
    \Gluon(174.36,104.26)(239.07,50.33){5.39}{6.86}
    \ArrowLine(303.78,50.33)(239.07,50.33)
    \ArrowLine(238.17,50.33)(238.17,104.26)\ArrowLine(239.97,50.33)(239.97,104.26)
\end{picture}
}
\end{center}
\caption{Feynman diagrams contributing to single-production of $t_R^*$ spin-3/2 resonances (double lines) via gluon-fusion and quark-antiquark annihilation.}
\label{feyn_single}
\end{figure}
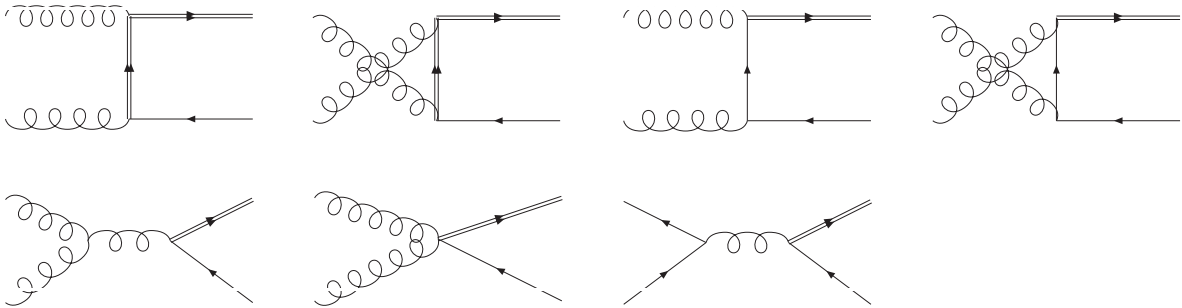

For single-production, the dominant diagrams come from the strong interaction in the operator of Eq.(\ref{chiral interactions}), and two types of diagrams need to be computed. 
The first type describes gluon-fusion into a rhd top and a $t_R^*$ resonance. There are six of these diagrams, as shown in Figure \ref{feyn_single}. The second type is annihilation 
of quark-antiquark pairs, again giving a rhd top and a resonance, also shown in Figure \ref{feyn_single}. There are also diagrams where 
a top and a gluon fuse to give a spin-3/2 resonance and a gluon, but the top content of the proton is miniscule and these diagrams give a negligible contribution.  The dominant process is gluon-gluon-fusion, as one would naively expect. 

We employed the package CalcHEP version 2.5.j \cite{Pukhov:2004ca} to calculate the cross-sections. We have computed analytically the decay width and the last two diagrams in Figure \ref{feyn_single} and checked their agreement with the results obtained using this package. The PDF used were the CTEQ6L set \cite{Pumplin:2002vw}, while $m_t= 171.2$ GeV \cite{Amsler:2008zzb}. 

\begin{figure}[htpb]
\begin{center}
\epsfig{file=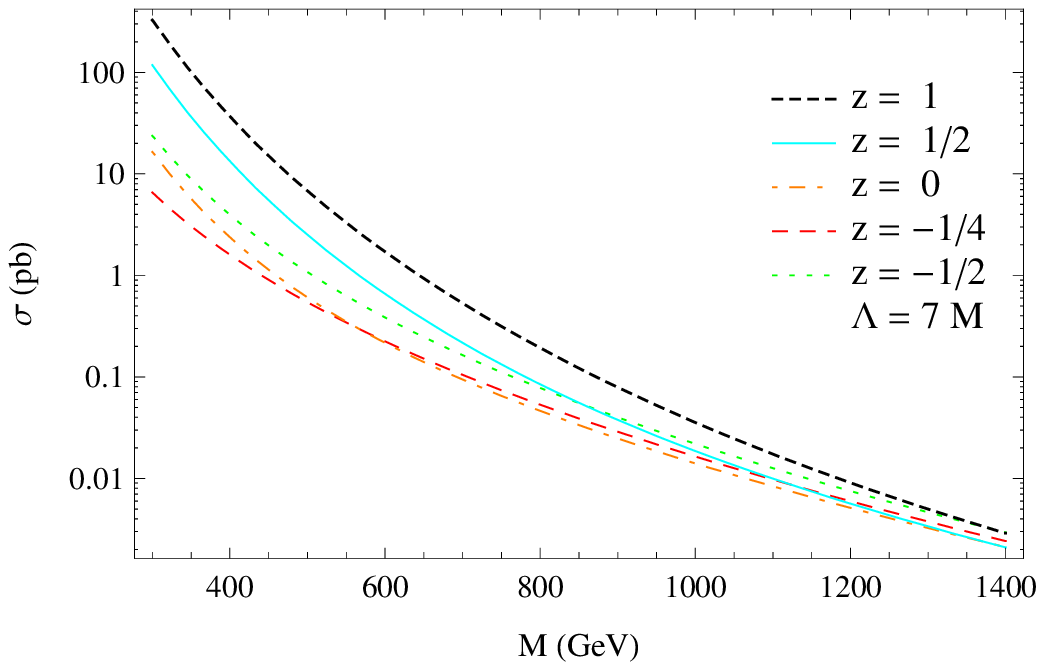, width=8.8cm} 
\epsfig{file=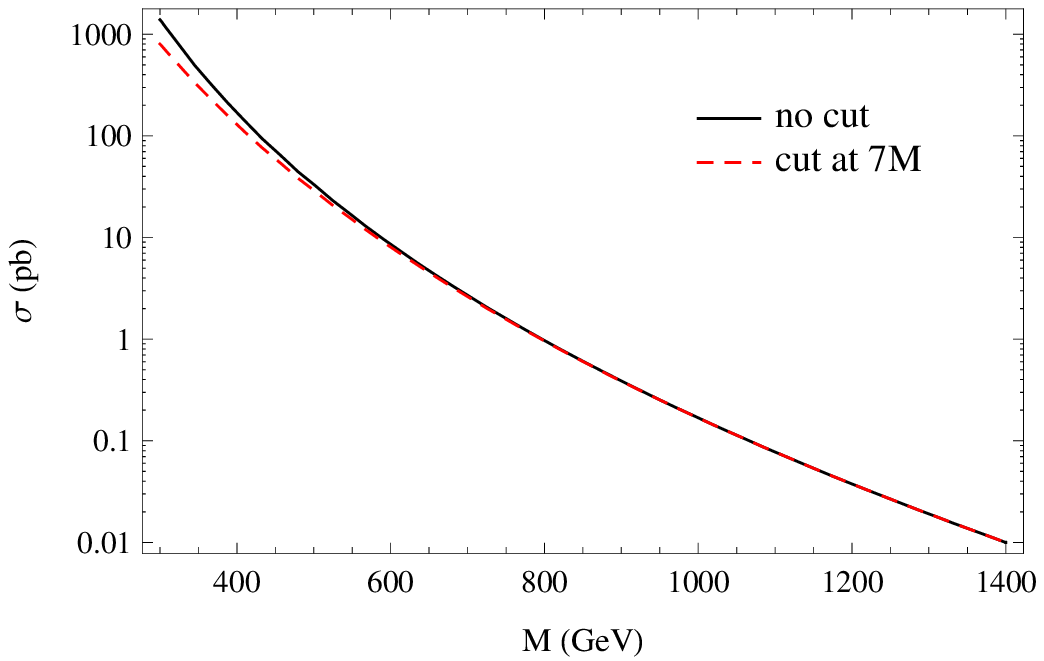, width=8.8cm} 
\caption[Resonance production cross-section as a function of its mass at the LHC.]
{Plots of the leading-order (LO) cross-section for production of $t_R^*$ against its mass, in $pp$ collisions at 14 TeV, with K-factor of 1. The figure on the left shows single-production for various values of  $z$. The plot on the right displays the total cross-section for pair-production, with and without a unitarity cut.} 
\label{cross-pair}
\end{center}
\end{figure}

In Figure \ref{cross-pair}, we show the behaviour of the hadronic cross-sections against the mass of the resonance. The plot on the left shows the single-production cross-section for various values 
of the parameter $z$. Guided by our estimate of the unitarity-violation scale, we use a value of $\Lambda = 7 M$, with smaller values of $\Lambda$ leading to a higher single-$t_R^*$ production rate as the latter goes like $1/\Lambda^2$. Recall that $\Lambda$ is a vestige of the five-dimensional theory, and will thus depend on the profiles of the fields in the extra-dimenion, and hence on the mass of the resonance. Note that the only diagrams which have a $z$-dependence are those with intermediate spin-3/2 propagation. 

The plot on the right in Figure \ref{cross-pair} shows pair-production with and without a unitarity cut $s_0$.   
It is clear that pair-production dominates over single-production for most values of $z$, unless the unitarity cut is very low. Moreover, pair-production depends very weakly on the off-shell parameter $z$ and $\Lambda$ (for $\Lambda > 2$ TeV) and is therefore more model-independent from a theoretical perspective, with a smaller SM background, which suggests that this channel might be more promising for detection purposes. Although single-production might have interesting experimental signatures, we relegate its discussion to future work and focus on pair-production in what follows. Note that the effect of the cut-off at $s_0=7 M$ is very mild, reducing the production cross-section by $15 \%$ for $M=300$ GeV but only by $3 \%$ for $M=400$ GeV, and is negligible for larger masses. This is related to the fortuitous cut-off enforced by the steeply falling PDF at scales below the unitarity bound for almost all resonance masses considered.

\begin{figure}
\begin{center}
\epsfig{file=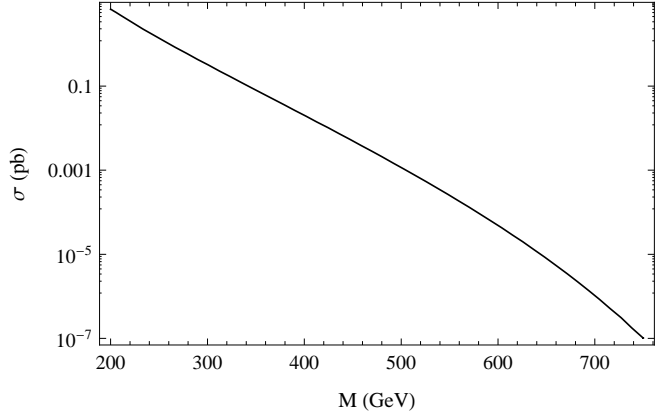, width=8.8cm} 
\caption[Resonance production cross-section as a function of its mass at Tevatron.]
{Plot of the LO cross-section for pair-production of $t_R^*$ against the their mass, in $p\bar{p}$ collisions at 1.96 TeV with K-factor of unity.} 
\label{tevatron}
\end{center}
\end{figure}

To our knowledge, no direct searches for spin-3/2 quark excitations have been done so far. Heavy up-like $SU(2)_W$ singlet spin-1/2 vector quark excitations have, 
however, been searched for in $p\bar p$ collisions by the CDF collaboration (\cite{Lister:2008is,Dobrescu:2009vz} and \cite{Pleier:2008ig}) at centre-of-mass energies of 1.96 TeV. 
These searches assume that the dominant decay channel of the spin-1/2 resonance is due to mass mixing with the SM top, decaying mainly into $Wd_{i}$, where $d_{i}$ is a 
down-type SM quark, and a lower-bound for the mass of such a quark excitation is quoted as 311 GeV. Calculating the cross-section for production of the $t_R^*$ spin-3/2 resonance at 
Tevatron energies (see Figure \ref{tevatron}), and using the SM background quoted in \cite{Lister:2008is,Dobrescu:2009vz}, we find a lower bound on its mass $M>340$ GeV for a 2$\sigma$ margin of exclusion. Note that this is pessimistically large, as it assumes $100\%$ decay into $Wd_{i}$, which we expect to compete with the other decay modes, as explained above. 
Interestingly, the existence of a mild excess of observed events at the Tevatron as reported by the CDF \cite{Lister:2008is} and D$\O$ \cite{D0note} collaborations 
can possibly be explained by a $t_R^*$ resonance with a mass of about 400 GeV. Because of the somewhat enhanced production cross-section of the $t_R^*$ at the Tevatron the excess is possibly better accommodated than with a standard top-prime, avoiding the need for the assistance of an extra $s$-channel colour-octet exotic \cite{Dobrescu:2009vz} (note that at the Tevatron the dominant production is via $q\bar q$ 
parton events, so the enhancement of the cross-section at the Tevatron is less than that at the LHC where $gg$ parton events occur with a $t-$ or $u$-channel $t_R^*$ propagator).

Finally, we now touch on the potential observability of these resonances at the LHC. We will content ourselves here with a rather simplistic measure of discovery, namely a simple signal vs background criterion $S/\sqrt{B} \geq 5$, with no regard to angular distributions, or the details of decay cascades.   All of the latter are clearly crucial facets of the analysis of the production and detection of any unstable higher-spin resonance, and they will be considered in upcoming work \cite{future}.

The $t_R^*\rightarrow W^+b$ decay mode is particularly interesting for direct searches, and we can in fact use it to estimate an upper-bound for the mass of spin-3/2 singlet resonances observable at the LHC. We utilize the background quoted in \cite{AguilarSaavedra:2005pv}, where pair-production of spin-1/2 up-like singlets decaying into $W^+b$ is considered. In \cite{AguilarSaavedra:2005pv}, it is found that spin-1/2 singlets with mass up to 1.1 TeV should be detectable at the LHC. Assuming all decays occur in the $W^+b$ channel and that the same K-factor as for spin-1/2 quark excitations applies, we find that spin-3/2 singlets with mass $\lesssim1.35$ TeV should be accessible at the LHC, reflecting the enhancement of the production rate for higher-spin particles. A combined search using the full set of decay channels including $t_R^*\rightarrow t g, tZ, tH$, and $t\gamma$ should raise this upper bound, so we view $M<1.35$ TeV as a conservative estimate for discovery of the $t_R^*$ at the LHC.  The merits of the various competing decay channels, as well as a more complete treatment of detection prospects, will be presented in forthcoming work \cite{future}.


\section{Conclusions}

In this work, we have considered the embedding of spin-3/2 Regge excitations of SM fields in Randall-Sundrum throats, which constitute a generic feature of any realization of these scenarios within string theory. We have considered the particular case where SM fermions and associated excitations are allowed to propagate in the bulk of the 5-dimensional geometry, in which case we find a reverse hierarchy amongst the KK masses of the (vector-like) spin-3/2 resonances, in the sense that the lightest states are those pertaining to the top in a large region of parameter space, with masses possibly below the TeV scale. This peculiar feature arises due to the way the fermions must be localized to account for the hierarchy in their masses, as explained above, and under the assumption of a flavour-blind string theory contribution to the bulk masses of these excitations. 

We expect top singlet excitations to decay via $t_R^*\rightarrow tg,\ W^+b,\  tZ,\ tH$ and $t\gamma$, whose model-dependent branching ratios may be comparable. Using the $ W^+b$ decay channel, we derive a lower bound of about 340 GeV on the mass of $t_R^*$ from Tevatron data, although this depends on the unknown branching ratio of this particular channel. Also, a reported excess of events in this channel can be possibly explained by a resonance with a mass of $\sim400$ GeV. If the $W^+b$ decay mode is dominant, we find that spin-3/2 singlets with mass up to 1.35 TeV should be discoverable at the LHC. 

The discovery of Regge excitations of SM fields may provide an unprecedented insight into the phenomenology of warped extra-dimensional models and string compactifications, and we are preparing a detailed analysis of the production of such states at the LHC, the associated decay cascades and the crucial issue of spin determination \cite{future}.

\vskip 1 cm

\centerline{\bf Acknowledgements}

\vskip 0.5 cm
The authors would like to thank Alan Barr, Alexander Shertsnev, Shamit Kachru, Jesse Thaler, Sandip Trivedi, and Stephen West for useful comments.  JMR thanks the UC Berkeley Center for Theoretical Physics for hospitality during the completion of this work. BH thanks Christ Church for financial support.  JGR is supported by FCT (Portugal) under the grant SFRH/BD/23036/2005. This work was partially supported by the EU FP6 Marie Curie Research and Training Network ``UniverseNet" (MRTN-CT-2006-035863), and by the STFC (UK).

\end{document}